\documentclass[twocolumn,trackchanges]{aastex63_2}

\newcommand{\dHybridR}{{\it dHybridR}}
\newcommand{\be}{\begin{equation}}
\newcommand{\ee}{\end{equation}}

\usepackage{amsmath}
\usepackage{graphicx}
\usepackage{dcolumn}
\usepackage{bm}
\usepackage{color}
\usepackage{xcolor}
\usepackage{ulem}

\submitjournal{ApJL}

\begin{document}

\preprint{}

\title{Dynamical Effects of Cosmic Rays on the Medium Surrounding Their Sources}

\correspondingauthor{Benedikt Schroer}
\email{benedikt.schroer@gssi.it}

\author[0000-0002-4273-9896]{Benedikt Schroer}
\affiliation{Gran Sasso Science Institute, Viale F. Crispi 7, 67100 L'Aquila, Italy}
\affiliation{INFN/Laboratori Nazionali del Gran Sasso, Via G. Acitelli 22, Assergi (AQ), Italy}

\author[0000-0002-7638-1706]{Oreste Pezzi}
\affiliation{Gran Sasso Science Institute, Viale F. Crispi 7, 67100 L'Aquila, Italy}
\affiliation{INFN/Laboratori Nazionali del Gran Sasso, Via G. Acitelli 22, Assergi (AQ), Italy}
\affiliation{Istituto per la Scienza e Tecnologia dei Plasmi, Consiglio Nazionale delle Ricerche, Via Amendola 122/D, 70126 Bari, Italy}

\author[0000-0003-0939-8775]{Damiano Caprioli}
\affiliation{Department of Astronomy and Astrophysics, The University of Chicago, 5640 S Ellis Ave, Chicago, IL 60637, USA}
\affiliation{Enrico Fermi Institute, The University of Chicago, 5640 S Ellis Ave, Chicago, IL 60637, USA}

\author[0000-0002-2160-7288]{Colby Haggerty}
\affiliation{Department of Astronomy and Astrophysics, The University of Chicago, 5640 S Ellis Ave, Chicago, IL 60637, USA}
\affiliation{Institute for Astronomy, University of Hawaii, 2680 Woodlawn Drive, Honolulu, HI 96822, USA}

\author[0000-0003-2480-599X]{Pasquale Blasi}
\affiliation{Gran Sasso Science Institute, Viale F. Crispi 7, 67100 L'Aquila, Italy}
\affiliation{INFN/Laboratori Nazionali del Gran Sasso, Via G. Acitelli 22, Assergi (AQ), Italy}





\date{\today}

\begin{abstract}
Cosmic rays (CRs) leave their sources mainly along the local magnetic field; in doing so they excite both resonant and nonresonant modes through streaming instabilities. The excitation of these modes leads to enhanced scattering and in turn to a large pressure gradient that causes the formation of bubbles of gas, CRs, and self-generated magnetic fields expanding into the interstellar medium. By means of hybrid Particle-In-Cell simulations, we show that, by exciting the nonresonant instability, CRs excavate a cavity around their source where the diffusivity is strongly suppressed. This finding invalidates the so far largely adopted flux-tube assumption, under which particles move along magnetic lines even in the nonlinear regime. This phenomenon is general and is expected to occur around any sufficiently powerful CR source in the Galaxy. Our results might provide a physical explanation of the numerous claims of suppressed CR diffusion around Galactic sources such as supernova remnants, pulsar wind nebulae, and stellar clusters.

\end{abstract}

\pacs{Valid PACS appear here}

\section{Introduction}
\begin{figure}
\begin{center}
\includegraphics[width=\columnwidth]{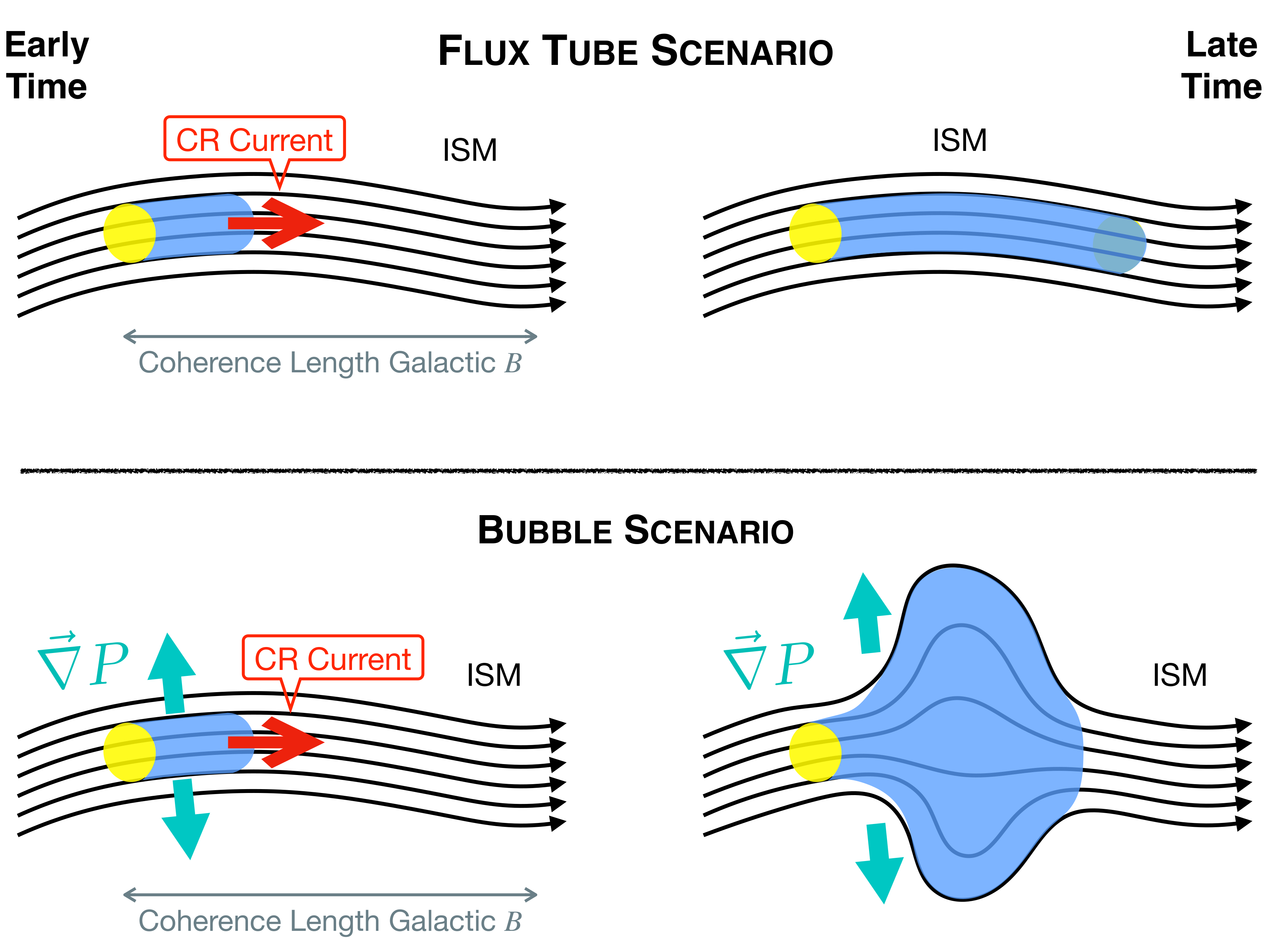}
\caption{Schematic view of the CR transport around a source in the standard flux tube assumption and as proposed here (bubble scenario).}
\label{fig:drawing}
\end{center}
\end{figure}

The transition from particles accelerated in astrophysical sources to CRs is still lacking a clear physical description despite its great importance for a variety of problems. The existing pictures of the CR injection into the interstellar medium (ISM) are rather simple: particles that have left their sources diffuse as test particles and their density eventually drops below the average CR Galactic density. 
The streaming of CRs getting away from a source along local Galactic magnetic field (GMF) lines
is responsible for the onset of resonant streaming instability \citep{Malkov,Dangelo,Nava,NavaRecchia}, which in turn reduces the diffusion coefficient and confines CRs in the surroundings of the source for a relatively long time. These conclusions are based upon the so-called flux tube assumption: CRs diffuse along the local GMF lines and excite a resonant streaming instability because of the gradients developed by their distribution function in that direction. As a consequence, a tube with a transverse size of the order of the geometric size of the acceleration region is filled by CRs even in the presence of self-generation, illustrated in the top panels of Figure \ref{fig:drawing}. Here we show that this assumption cannot be fulfilled in the region around a typical CR source.

The first point to notice is that the diffusion length of CRs that left the acceleration region with large enough energy exceeds the coherence scale of the GMF. Hence, their transport is not strictly diffusive, at least in the beginning, and these particles behave as a current beam, capable of exciting a nonresonant streaming instability \citep[NRSI;][]{Bell2004,AmatoBlasi09}. The larger growth rate of the NRSI helps in overcoming some of the limitations typically attributed to the resonant instability, e.g., turbulent damping \citep{Farmer_2004}.
On the same time-scale, particles are forced into a diffusive regime, with a diffusion coefficient that can be estimated to be much smaller than the one usually inferred on Galactic scales from secondary-to-primary ratios in the ISM. 

We demonstrate here, first by using analytic arguments, that 1) streaming particles are able to excite fast-growing modes through NRSI, and 2) CRs self-confine themselves in a region surrounding the source. By doing so, CRs establish a pressure gradient with respect to the region outside the flux tube. This configuration is unstable since the pressure gradient in the transverse direction causes an expansion of the tube and a partial evacuation of the surrounding region, as illustrated in the bottom panels of Figure \ref{fig:drawing}.
We confirm this expectation by using two-dimensional (2D) hybrid (kinetic protons, fluid electrons) Particle-In-Cell (PIC) simulations \citep{2019ApJ...887..165H} describing the behavior of CRs leaving their source. The coupling of CRs and the background plasma has been considered in several numerical works, within a magnetohydrodynamics (MHD) \citep{2012ApJ...761..185Y,2016A&A...585A.138D,2017MNRAS.465.4500P,2017ApJ...834..208R,Jiang_2018,2019MNRAS.489..205W,2019A&A...631A.121D,10.1093/mnras/stab142,10.1093/mnras/staa2025}, a combined MHD-kinetic \citep{bai+15,2018MNRAS.476.2779L,vanmarle+19} or a hybrid kinetic \citep{2019ApJ...887..165H,2020ApJ...905....1H,2020ApJ...905....2C} framework.
Although a fluid approach allows to describe larger systems, here we focus on a kinetic treatment since it guarantees that the microphysics processes responsible for the bubble formation are retained.
We use a 2D domain initialized with a uniform magnetic field and a localized CR source; CRs that have left the source and move in the ISM self-consistently trigger the NRSI and start scattering, creating an overpressure that inflates an extended bubble around the source. Our setup is innovative, in that it allows us to capture the transverse expansion of the bubble that is inevitably lost in the simple existing nonlinear models \citep{Malkov,Dangelo,Nava,NavaRecchia}. It is worth stressing that the bubble formation due to the pressure gradient is generic and should be present also when the streaming instability is limited to the resonant regime.

We find that the magnetic field inside the bubble is due both to the excitation of NRSI and to turbulent motions induced by the expanding plasma bubble. This finding is of great importance for several reasons: first, the CR confinement in the region around the sources may lead to an excess grammage that might affect the observed secondary-to-primary ratios (such as B/C) and the production of positrons \cite[]{2010PhRvDCowsik,2018APhLipari}; second, the CR overdensity around sources, caused by the reduced diffusivity, may produce an excess of $\gamma$-ray production possibly connected with recent observations from regions around selected supernova remnants \cite[]{w28}, star clusters \cite[]{Felix} and PWNe \cite[]{hawc}. 

In Sect. \ref{sect:theory}, we first discuss the development of the NRSI induced by CRs leaving their sources and the conditions that are expected around such sources. Then, in Sect. \ref{sect:PIC}, we introduce the kinetic simulations used to model the motion of CRs in the ISM after they have left the sources. Our main results and their implications are discussed in Sect. \ref{sect:results}, while we conclude in Sect. \ref{sect:concl}.

\section{Streaming instability}
\label{sect:theory}
In this section we present a back-of-the-envelope calculation to obtain a first intuition for the processes triggered by CRs around their sources. For the sake of clarity we assume that the source is a SNR where particles are energized through diffusive shock acceleration. The particle leakage at the highest energy reached at any given time $t$ during the ST phase is a crucial ingredient of the mechanism and the very reason why scattering occurs due to the excitation of NRSI by the same escaping particles \cite[]{Bell2004,Schure2014MNRAS}. Although the number density of such particles is small, their current can be demonstrated to be the same as at the shock \cite[]{Bell2013MNRAS}. The particles that left the acceleration region are the main characters in the picture discussed below.

For the sole purpose of providing simple estimates of the phenomena that we are interested in, we assume that the diffusion coefficient in the Galaxy is of order $D(E)\simeq 3\times 10^{28}E_{GeV}^{1/2}$, for $E \gtrsim 10$ GeV/n. More refined treatments of Galactic diffusive transport return better descriptions of its normalization and energy dependence \citep[e.g.,][]{CarmeloNuclei,Evoli2020}. 
In terms of diffusion length $\lambda$, this reads $D(E)= v \lambda(E)/3$, and, assuming $v\approx c$ (being $c$ the speed of light), $\lambda(E)\simeq 1~{\rm pc}~E_{GeV}^{1/2}$. If the coherence scale of the GMF is taken to be $l_{c}\sim 50~{\rm pc}$, the diffusion length exceeds $l_c$ at $E\gtrsim 2.5$ TeV. At these energies particle transport immediately around the sources is roughly ballistic, and a coherence scale is covered in a time $\tau_{c}\simeq 3 l_{c}/c\sim 500$yr. The corresponding density of CRs can be estimated as
\be
n_{CR}(>E)\approx \left(\frac{E_{s}}{E}\right) \frac{\xi_{CR}}{\pi R_{s}^{2} c T_{s} \Lambda},  
\ee
where we assumed that the differential spectrum of particles is $\sim E^{-2}$ and $\Lambda=\ln(E_{max}/mc)\sim 15$. Here $R_{s}$ and $T_{s}$ are the source size and the typical time scale in which the energy $E_{s}$ is released. For a SNR, we choose $T_{s}$ of order the beginning of the Sedov phase ($\sim$300 yr) and $R_{s}$ as the corresponding SNR size ($\sim3$ pc). The typical energy released in the form of kinetic energy of the expanding gas is assumed to be $E_{s}=10^{51}$ erg, adequate for most SNRs, while a typical conversion efficiency to CRs of $\xi_{CR}=0.1$ is chosen \citep{caprioli+14a}. 
For these values, one infers a CR energy density $n_{CR}(>E) E\sim 54 ~\rm eV/cm^{3}$, substantially larger than the typical GMF energy density.
It is easy to see that this holds true also for CRs released after the end of the Sedov phase and propagating diffusively in the source surroundings.
Even with observationally-preferred \citep{2019PhRvD..99j3023E} and  theoretically-motivated \citep{2020ApJ...905....1H,2020ApJ...905....2C} steeper spectra, e.g., $E^{-2.3}$, the energy density is decreased only by a factor of $2$ at $2.5\,$TeV energies, which still significantly exceeds the energy density of the GMF, $\sim 0.2 ~\rm eV/cm^{3}$ for $B_0=3\mu$G. 
When the CR momentum flux (related to the CR energy density) exceeds the local magnetic pressure, a nonresonant hybrid instability is excited \cite[]{Bell2004}.
The fastest-growing mode has a growth rate $\gamma_{max}=k_{max}v_{A}$ where 
\begin{equation}
k_{max}B_{0}=\frac{4\pi}{c}J_{CR}(>E) \, ,
\end{equation}
the Alfv\'en speed $v_A = B_0/\sqrt{4\pi m_i n_0}$, with $n_0$ the number density of the ISM and $m_i$ the ion mass
and the electric current of particles with energy $\gtrsim E$: $J_{CR}(>E)=e n_{CR}(>E) c$.
Using the previous estimates, these modes 
grow on a time scale $\gamma_{max}^{-1} \approx 1.1 \rm (E/2.5 TeV)$ yr. 
Moreover, $k_{max}^{-1}$ is smaller than the gyroradius of the particles dominating the current, 
hence the current is not affected by the growth, at first instance. 
On longer time scales the magnetic field
is expected to saturate when the energy density in the amplified field becomes comparable with 
the one of escaping CRs. This condition is 
equivalent to assuming that power is transferred from small scales $k_{max}^{-1}$ to the scale of the particles' gyroradius in the amplified magnetic field \citep{Bell2004, caprioli+14b}. 
For the above parameters, we expect at most $\delta B/B_{0}$ of a few. However this field has a typical scale of the order of the gyroradius of the CRs dominating the current. For 2.5 TeV particles this corresponds to $\sim$ few $10^{-4}$ pc, much smaller than $l_{c}$. Hence, the effects on the diffusion properties of CRs are dramatic, despite the fact that $\delta B/B$ is not very large. For a spectrum $E^{-2}$ we expect diffusion to occur at roughly the Bohm rate \citep[e.g.,][]{reville+13,caprioli+14c},  
meaning that CRs with energy 2.5 TeV would have a diffusion coefficient $D(E)= r_{L} c/3\approx 10^{25}\rm cm^{2}s^{-1}$, several orders of magnitude smaller than expected at the same energy for Galactic CRs. 

This simple estimate shows that it is impossible for high-energy CRs to travel in the ISM surrounding a source ballistically, since their streaming induces the excitation of modes that scatter particles, thereby confining them in the circumsource region.
It follows that their density and local pressure increase by a large amount because of the effect of diffusion.
This physical picture, that we specialized to the case of ballistic streaming of CRs, is qualitatively identical to that investigated in previous literature as due to resonant streaming instability \citep{Malkov,BlasiDangelo,Nava,NavaRecchia}. 
However, all these approaches missed a crucial piece of information, which we introduce here:
the large overpressure in the tube of cross section $\sim \pi R_{s}^{2}$ with respect to the external ISM leads to a pressure gradient in the transverse direction that is expected to result in the expansion of the tube, so that one of the main assumptions of previous calculations gets invalidated.
This expectation has several important implications: 
1) since the gas, magnetic field and CRs are initially in pressure balance in the Galactic ISM,
the injection of new nonthermal particles near a source is bound to break such balance and lead to excitation of the NRSI; the accumulation of CRs in the near source region unavoidably leads to the creation of an overpressurized region expanding into the surrounding ISM. 2) The role of streaming instability is only that of bootstrapping the process by making particles scatter, thereby enhancing the local pressure. At that point, the bubble expansion, driven by the CR overpressure, starts and likely continues until pressure balance is achieved or damping of the turbulence leads to particle escape. In the first case, the final size of the CR bubble is estimated to be about $L\approx (\xi_{CR} E_{s}/\Lambda P_{ISM})^{1/3} \sim 60\,$pc, where $P_{ISM}$ is the pressure of the ISM.
We will discuss these points in more detail below, based on the results of the PIC simulation and having in mind observational evidence of reduced CR diffusivity based on $\gamma$-ray observations.

\section{PIC simulations}
\label{sect:PIC}
To study the nonlinear coupling of 
streaming CRs and the background plasma, we perform self-consistent simulations using \dHybridR{}, a relativistic hybrid code with kinetic ions and (massless, charge-neutralizing) fluid electrons \citep{2007CoPhC.176..419G,2019ApJ...887..165H}. We consider here an adiabatic closure for the electron pressure, i.e., $P_e\propto \rho^{5/3}$ \citep{2018JPlPh..84c7101C, 2020ApJ...905....1H}.
\dHybridR{} is the relativistic version of the Newtonian code {\it dHybrid} \cite[]{2007CoPhC.176..419G} and it is capable of properly simulating CR-driven streaming instabilities \citep{2019ICRC...36..279H, 2019ICRC...36..483Z}.
Hybrid codes are better suited to self-consistently simulate the long-term, large-scale coupling of CRs and background plasma than fully-kinetic PIC codes since they do not need to resolve small electron scales, usually dynamically negligible.

In simulations, physical quantities are normalized to the number density ($n_0$) and magnetic field strength ($B_0$) of the initial background plasma. Lengths, time and velocities are respectively normalized to the ion inertial length $d_i = v_A/\Omega_{ci}$, the inverse ion cyclotron frequency $\Omega_{ci}^{-1}$, and 
$v_A$. The background ion temperature is chosen such that $\beta_i = 2 v_{th,i}^2/v_A^2=2$, i.e. thermal ions gyroradius $r_{g,i}=d_i$. The system is 2D ($x-y$) in physical space and retains all three components of the momenta and electromagnetic fields. We discretized the simulation grid, of size $5000\times 7000\,d_i$, with $7500\times10500$ cells (i.e. $\Delta x=\Delta y \simeq 0.66 d_i$). Open boundary conditions are imposed in each direction for the CRs and on $x$ for the background plasma; the $y$ direction is periodic for thermal particles.
A background magnetic field, directed along $x$ and of strength $B_0$, is embedded in the simulation domain. 
The time step is $0.01\,\Omega_{ci}^{-1}$ while $c=20\,v_A$. The background plasma, described with $N_{\rm ppc}=4$ particles per cell, has density $n_0$ and its distribution is Maxwellian. CRs, discretized with $N_{\rm ppc}=16$, are injected at the left boundary at $x=0$ in a small stripe $3200d_i<y<3800 d_i$ with an isotropic momentum distribution with $p_{\text{total}}=100\,mv_A$, i.e. Lorentz factor $\gamma\approx 5$, and $n_{\text{CR}}=0.0133\,n_0$. CR injection is continuous in time and numerical parameters are comparable with the SNR case, in the sense that the ratio of energy densities of CR particles and thermal energy in the beginning is $\sim (n_{\text{CR}}/n_0)(c/v_A)^2\gamma\sim 26.$

\begin{figure}
\begin{center}
\includegraphics[width=0.48\textwidth]{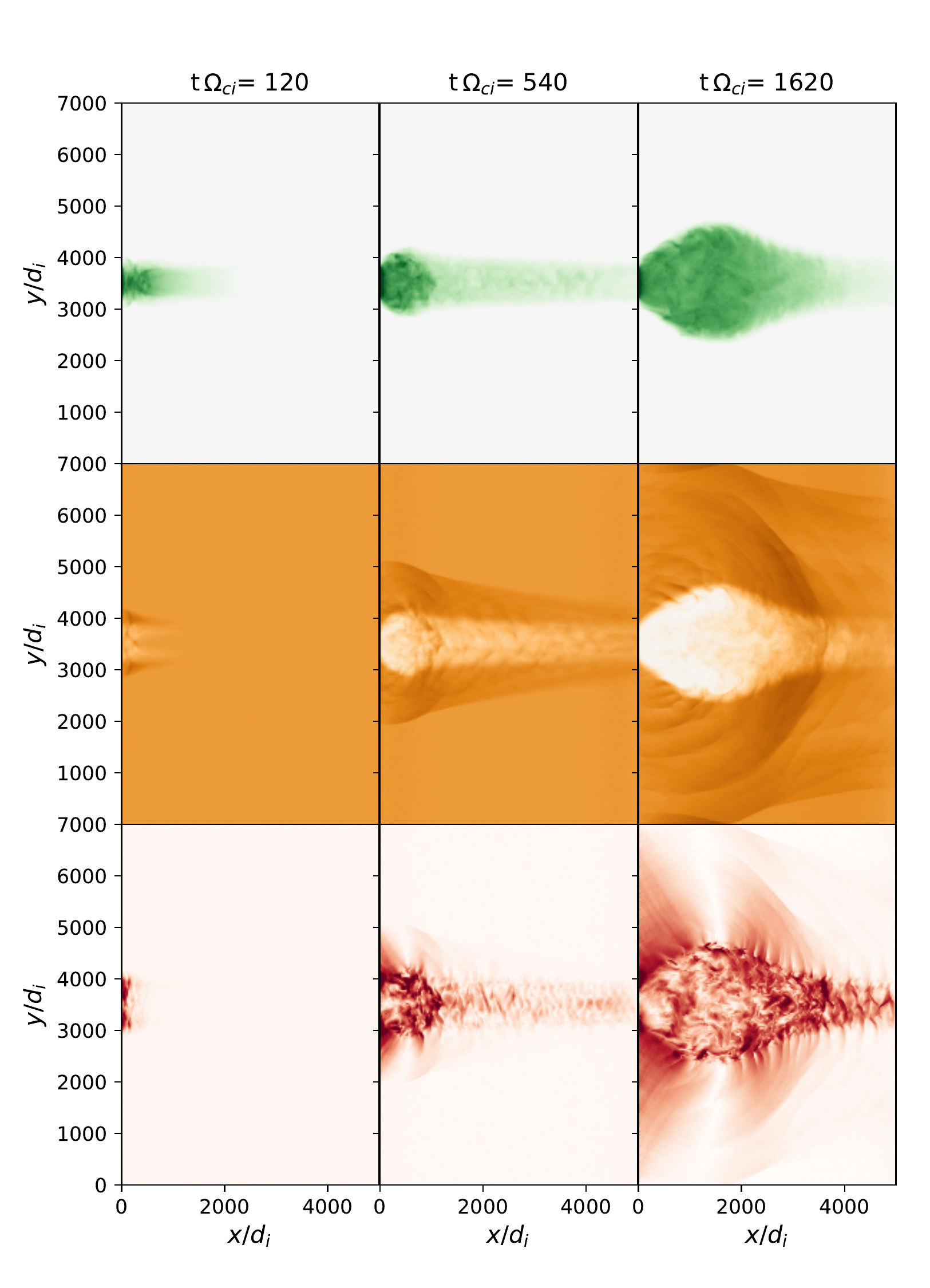}
\caption{Contour plots of CR density (top), background plasma density (center), and perpendicular magnetic field (bottom) at three times in the simulation.
An animated version of this figure is available in the HTML version of the article showing the time evolution of these quantities throughout the whole simulation and demonstrating the growth of the bubble.}
\label{fig:snaps}
\end{center}
\end{figure}

\section{Results}
\label{sect:results}
As discussed above, the excitation of streaming instability acts as a bootstrapping process for seeding the overpressurized region around the source. Although this may be expected to occur even due to resonant streaming instability, we checked that both right- and left-handed waves are produced, as expected if the NRSI is also excited.

In principle, this configuration may also produce other instabilities, e.g., driven by pressure anisotropies \citep{2020ApJ...890...67Z}.
Once the particles start scattering on these self-generated perturbations, they move slower in the $x$-direction and their number density increases. 
This can be seen in the top panels of Figure \ref{fig:snaps}, which display the CR density $n_{CR}$ at three different times in the simulation.
Several interesting aspects arise from this figure: 
first, at early times, CR presence is limited to a small region around the injection location, and the region occupied by CRs has basically the same transverse size as the source itself (in fact somewhat larger because their injection is isotropic, hence CRs are initially distributed on a region exceeding the source size by a gyroradius on both sides). 
Particles are still streaming ballistically in the $x$-direction. At later times, the density of CRs around the source increases and the region filled by CRs expands in the transverse direction as a result of the overpressurization of the flux tube due to scattering. This finding invalidates the assumption, made in previous work on CR transport around sources, that CRs propagate in a tube with the same transverse size as the source (see also Figure \ref{fig:drawing}).

The force associated with the gradient of CR pressure in the perpendicular direction causes a partial evacuation of the plasma previously located inside the bubble, as can be seen in the central panels of Figure \ref{fig:snaps} (gas density, $n_{gas}$). 
While the bubble expands, the gas density in the center of the bubble decreases while the gas density on the outskirts of the bubble increases and density waves are launched outward in the simulation box. In fact we stop the simulation when those waves reach the boundary in the $y$-direction.
The bubble expansion triggered by CR scattering is due to the generation of magnetic perturbations in the directions perpendicular to the initial background magnetic field, which are initially absent. This is illustrated in the last row of plots of Figure \ref{fig:snaps}, where we show $B_{\perp}$ at three different times. At early times there is no turbulent magnetic field. The streaming of particles along the $x$-direction drives the formation of a highly-structured $B_{\perp}$. 
The self-generated magnetic field follows the expansion of the bubble and determines the local rate of particle scattering in the whole volume filled by CRs. 
The magnetic field seems particularly strong on the edges of the bubble, signaling that the bubble is wrapped in an envelope of swept up compressed field lines. Inside the bubble the field is irregular, as it should be if responsible for CR scattering. 

\begin{figure}
\begin{center}
\includegraphics[width=0.49\textwidth]{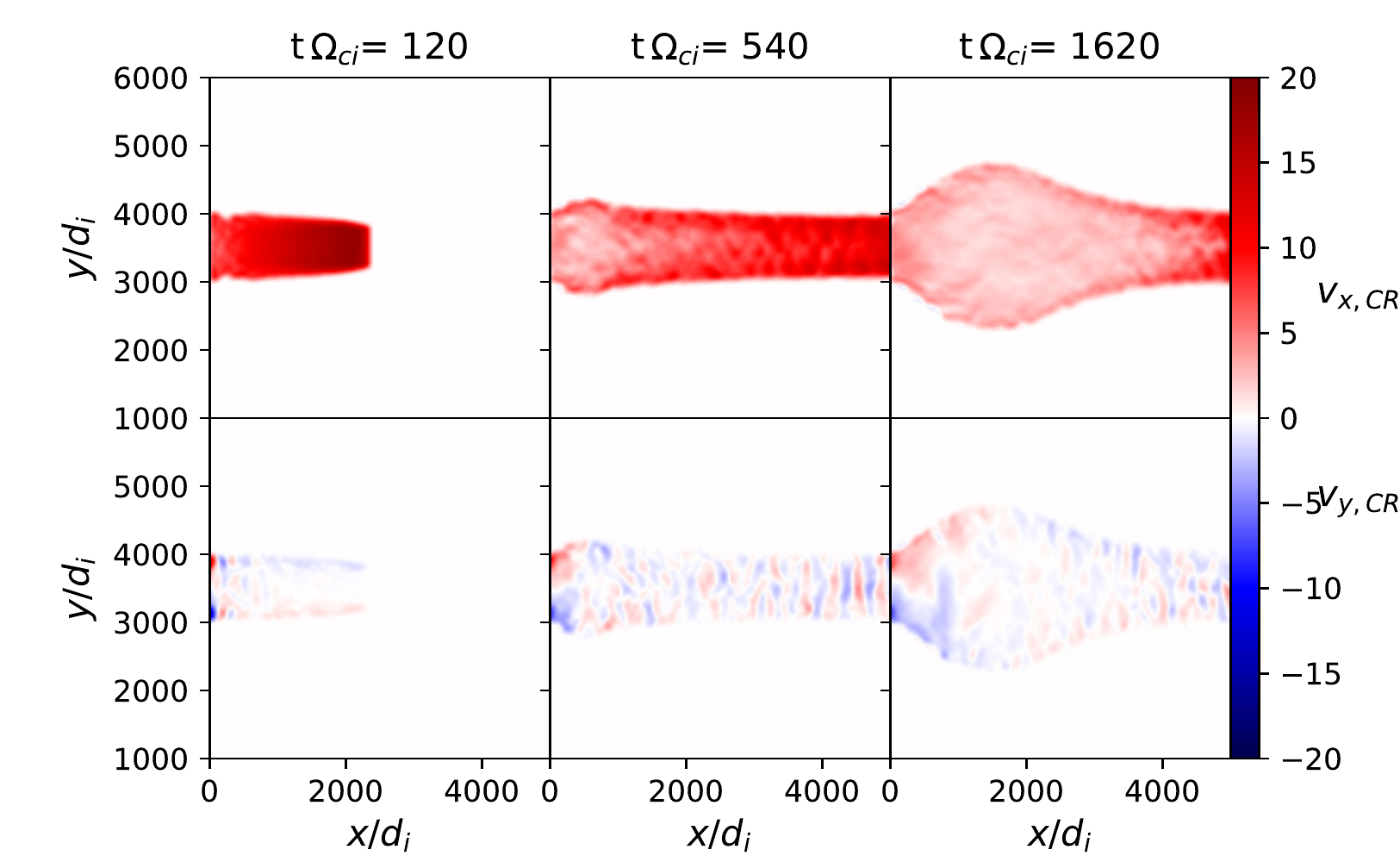}
\caption{Contour plots of the CR bulk speed along $x$ (top) and $y$ (bottom) at three instants of time.
}
\label{fig:velocity}
\end{center}
\end{figure}

The fact that CR transport around the source gets profoundly affected by this turbulent field is proven by different pieces of evidence. One is illustrated in Figure \ref{fig:velocity}, which shows the contour plots of the $x$ (top) and $y$ (bottom) components of the mean CR speed in the simulation domain.
At early times the velocity component $v_{x}$ is strongly directed toward the positive $x$-direction.
At later times, the CR scattering off self-generated waves leads to isotropization in that direction, so that inside the bubble the mean $v_{x}$ tends to vanish, a clear sign of particle diffusion. Since particles are injected isotropically in the half-sphere $v_{x}>0$, the mean of the perpendicular component $v_{y}$ is vanishingly small at all times. 

We have roughly estimated the parallel diffusion coefficient by taking the ratio of the square of the bubble size in the $x$-direction at a given time and time itself. With this rather simple estimate we found that the diffusion coefficient is a few times the Bohm diffusion coefficient. This further proves that there is strong scattering induced by self-produced magnetic field perturbations.

\section{Conclusions}
\label{sect:concl}
We demonstrated that for typical sources of CRs in the Galaxy the ballistic propagation of TeV CRs around the sources is inhibited by self-generation of magnetic perturbations, which in turn lead to CR scattering and to the creation of an overpressurized region. The pressure gradient with respect to the external ISM produces the expansion of this bubble and the evacuation of the region. 
We expect that all typical CR sources should be surrounded by extended regions of reduced diffusivity and local overdensity of CR particles. 
It is worth stressing that limitations associated with the finite size of the simulation box force us to limit the evolution of these phenomena to relatively short times compared with those of relevance for astrophysical situations. 
It follows that some degree of inference needs to be adopted to extrapolate our conclusions to more realistic situations. 
For instance, it is likely that  CR-excavated bubbles keep expanding under the action of the pressure gradient even after CR injection by the source turns off, a scenario too expensive to simulate given the long time scales involved.
On the other hand, our findings are completely consistent with what we could expect based on simple arguments and illustrated in Figure \ref{fig:drawing}. Hence, we dare to infer that the bubble of reduced diffusivity inflated by CR in the region around sources must be a generic phenomenon and is likely to be related to the claims of small diffusion coefficients based upon $\gamma$-ray observations from molecular clouds in the regions around SNRs \citep{w28,Gabici2010} and star clusters \citep{Felix}. 
Indeed, the expected size of the bubble, $\sim 50-100\,$pc if pressure balance is assumed, is in agreement with these observations. The case of $\gamma$-ray emission from the regions surrounding pulsar wind nebulae \citep{hawc} can be qualitatively different in that the escaping particles are electron-positron pairs for which, at zeroth order, the absence of a net current inhibits the onset of Bell instability. This will be discussed in a forthcoming article. 
The role of resonant streaming instability was investigated by \cite{2018PhRvD..98f3017E}, who concluded that in typical conditions it is not sufficient to explain HAWC observations.

The very existence of these bubbles has many implications that bear the potential to shake the pillars of CR transport: First, depending on the details of the gas density and presence of clouds in the circumsource regions, an appreciable grammage could be accumulated by CRs due to the reduced diffusivity, thereby affecting the secondary-to-primary ratios measured at the Earth \cite[]{2010PhRvDCowsik,2018APhLipari}. 
Second, electrons leaving the source diffuse through the bubble with small diffusion coefficient, in times that may become comparable with those associated with radiative losses, thereby producing a possibly pronounced difference between the spectrum of protons (or nuclei) and that of electrons \citep{diesing+19,2021arXiv210302375C}. 

\acknowledgments
Simulations were performed on computational resources provided by the University of Chicago Research Computing Center, the NASA High-End Computing Program through the NASA Advanced Supercomputing Division at Ames Research Center, and XSEDE TACC (TG-AST180008). 
DC was partially supported by NASA (grants NNX17AG30G, 80NSSC18K1218, 80NSSC20K1273, and 80NSSC18K1726) and by NSF (grants AST-1714658, AST-2009326, AST-1909778, PHY-1748958, and PHY-2010240).

\end{document}